\documentclass[superscriptaddress, aps, prx, longbibliography]{revtex4-2}
\usepackage{graphicx}
\usepackage{array}
\usepackage{siunitx}
\usepackage[section]{placeins}
\usepackage{hyperref}
\usepackage{natbib}

\newcommand{\FOM}{\mathrm{FOM}_{D_2}}

\begin{document}

\title{Current experimental upper bounds on spacetime diffusion}
\date{February 2024}

\author{Martijn Janse}  \email{These authors share first authorship and contributed equally.}
\author{Dennis G. Uitenbroek} \email{These authors share first authorship and contributed equally.}
\author{Loek van Everdingen} \email{These authors share first authorship and contributed equally.}
\author{Jaimy Plugge} 
\author{Bas Hensen} \email{hensen@physics.leidenuniv.nl}
\author{Tjerk H. Oosterkamp} \email{oosterkamp@physics.leidenuniv.nl}

\address{Leiden Institute of Physics, Leiden University, P.O. Box 9504, 2300 RA Leiden, The Netherlands}

\begin{abstract}

A consistent theory describing the dynamics of quantum systems interacting on a classical spacetime was recently put forward by Oppenheim et al..[1, 2]. Quantum states may retain their coherence, at the cost of some amount of stochasticity of the spacetime metric, characterized by a spacetime diffusion parameter. Here, we report existing experimental upper bounds on such spacetime diffusion, based on a review of several types of experiments with very low force noise over a broad range of test masses from single atoms to several kilograms. We find an upper bound at least 15 orders of magnitude lower as compared to the initial bounds for explicit models presented by Oppenheimn et al. The results presented here provide a path forward for future experiments that can help evaluate classical-quantum theories.
\end{abstract}

\maketitle

\section{Introduction}
Oppenheim et al. have pointed out that quantum mechanics can be combined with a classical interpretation of gravity while still providing a self-consistent framework~\cite{OppenheimNatCom, OppenheimPRX}. To this end, they have introduced a spacetime diffusion parameter, $D_2$, which describes the fluctuations arising due to the interplay between a quantum system and its surrounding classical gravitational potential. Put differently, this is the rate of diffusion of the conjugate momenta of the Newtonian potential, that is necessary to preserve any amount of coherence of the quantum system subject to classical gravity.\\

Three different models were put forward in Oppenheim et al.~\cite{OppenheimNatCom} to describe this spacetime diffusion parameter, each with their own upper and lower bound. The ultra-local continuous model has already been ruled out in the original paper by comparison with experiments. However, the remaining two models, i.e. the ultra-local discrete and the non-local continuous model, are not yet rejected (Equations 46 and 47 respectively in~\cite{OppenheimNatCom}). In the ultra-local discrete model $D_2$ has an upper bound of

\begin{equation}
   \frac{l_P^3}{m_P} D_2 \leq \frac{\sigma_a^2  N \Delta T r^4_N}{m_NG^2}
   \label{discrete_model}
\end{equation}
 and in the non-local continuous model $D_2$ has an upper bound of
\begin{equation}
   l_P^2 D_2 \leq \frac{\sigma_a^2  N \Delta T r^3_N}{G^2}.
   \label{cont_model}
\end{equation}
Here, $l_P$ is the Planck length, $m_P$ the Planck mass, $G$ the gravitational constant, $r_N$ the radius of the nucleus and $m_N$ its mass. In addition to this upper bound, Oppenheim et al. also introduce a lower bound based on decoherence experiments. We do not address this lower bound extensively in this work. In order for the theory to be ruled out in the current formulation, there are still 26 and 24 orders of magnitude of $D_2$ to be overcome in the ultra-local discrete model and the non-local continuous model respectively.\\

Leaving the natural constants aside and assuming the radius and mass of the nucleus to be constant in the experiment we consider, we find three parameters that combine in the same way for both expressions. These are the acceleration noise $\sigma_a$, the number of nuclei $N$, and the measurement time $\Delta T$. These three parameters essentially introduce a figure of merit $\FOM$, which needs to be minimized in order for the upper bound on $D_2$ to be lowered for both models. The $\FOM$ may be deduced from precision experiments that minimize acceleration and/or force noise

\begin{equation}
    \FOM = \sigma_a^2  N \Delta T = S_aN.\\
\end{equation}

Here, we use the spectral density $S_a = \sigma_a^2 \Delta T$. To facilitate the comparison with different experiments, we prefer to use the spectral densities $S_a$ rather than the variance $\sigma_a^2$, such that the measurement time $\Delta T$ drops out of the equations. Since many experiments focus on the force noise, $S_F$, rather than the acceleration noise, it is necessary to convert $S_a$ to $S_F$, using $\sigma_F = m\sigma_a $ or $S_F=m^2S_a$. The mass $m = M N/N_A $ is related to the number of particles $N$ through the average molar mass $M$ of the materials used and Avogadro's number $N_A$. In the case of molecules, we multiply the number of particles $N$ by the number of nuclei per molecule.\\

In this work we point out that the figure of merit for modern-day experiments, ranging from atom interferometry to the LISA Pathfinder mission, can vastly outperform the figure of merit of the Cavendish torsion balance experiment which Oppenheim et al. used as a first source of experimental bound on their models~\cite{OppenheimNatCom}. For this first bound, Oppenheim et al. used $N=\SI{e{26}}{}$, $\sigma_a=\SI{e{-7}}{m/s^2}$ and $\Delta T=\SI{100}{s}$, leading to $\FOM=\SI{e{14}}{m^2/s^3}$.\\

\section{Results}
\begin{figure}
    \centering
    \includegraphics[width = \textwidth]{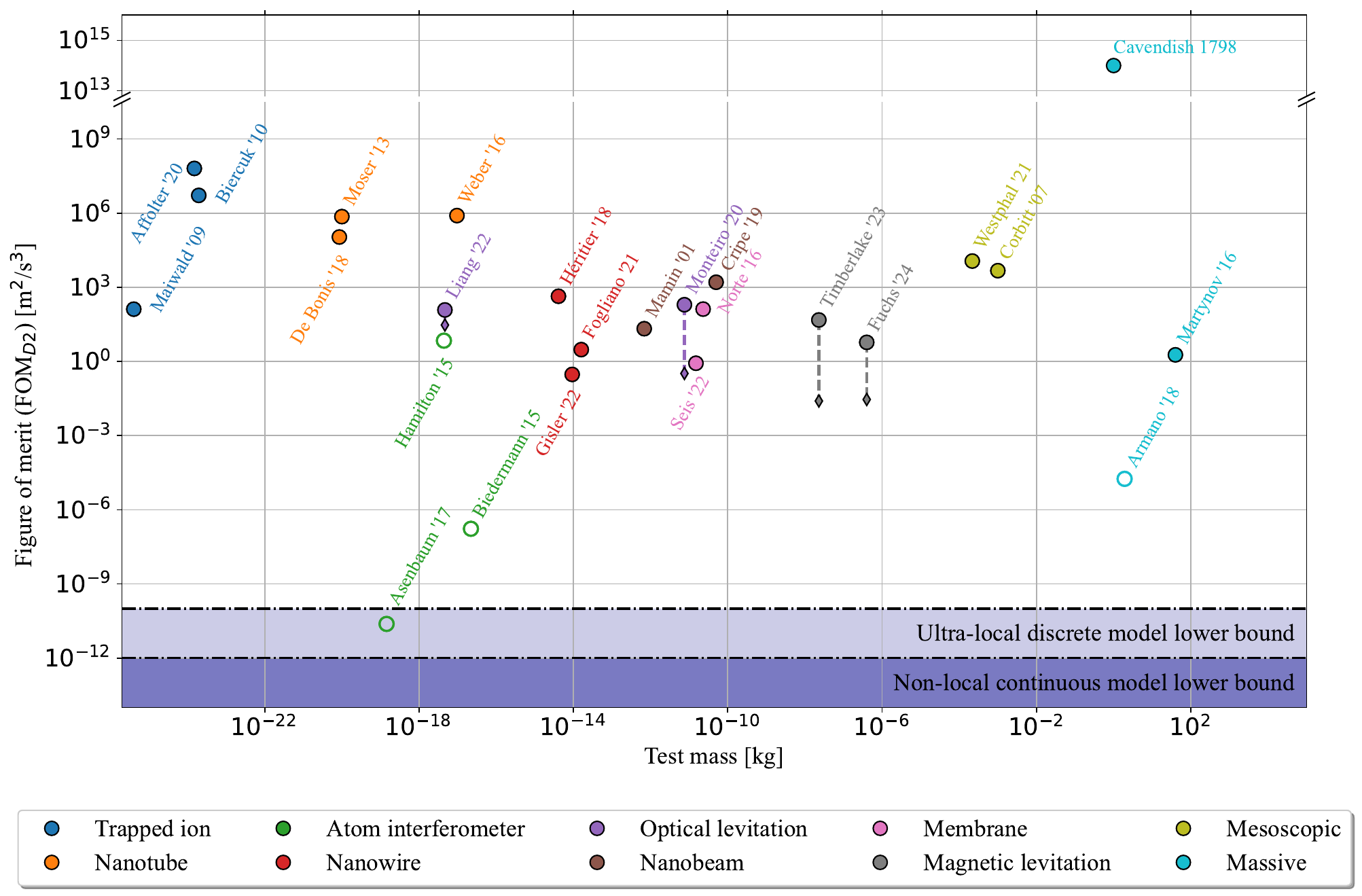}
    \caption{The figure of merit $\FOM$ as calculated for different kind of force sensitivity experiments. Different types of experiments along a broad mass range were considered. Plotted experiments are considered to be representative of their type. The actual measurement reported for each experiment is plotted as a circle. A diamond, indicating a potential thermal force limit, is added to experiments in which the total force noise is at least two times larger than the thermal force noise. Markers with a colored edge indicate that the experiment was a differential measurement and therefore the validity of the calculated $\FOM$ is questioned. The shaded areas in the lower part of the plot indicate the lower bound that was put on $\FOM$ by Oppenheim et al. for respectively the ultra-local discrete model and the non-local continuous model.}
    \label{Figure1}
\end{figure}

In Figure \ref{Figure1} we show $\FOM$ as a function of the test mass for different types of experiments with low force or acceleration noise. The underlying data is shown in Table I in Appendix A. We observe that experiments with test masses in the range between $\SI{e{-22}}{kg}$ and $\SI{e2}{kg}$ can yield a stronger upper bound compared to the Cavendish torsion balance experiment. An experiment by Gisler et al.~\cite{Gisler} achieves a $\FOM$ that is 15 orders of magnitude lower by using a nanowire mechanical resonator with an effective mass of $\SI{9.3e{-15}}{kg}$. This measurement results in the lowest $\FOM$ with an absolute measurement of acceleration noise on Earth. A slightly higher $\FOM$ is achieved by Martynov et al.~\cite{Martynov}, which concerns the LIGO mirror with a test mass of several kilograms. This is an indication that the minimal achievable acceleration noise is largely independent of the test mass.\\

On top of that, we also observe an improvement of 19 orders of magnitude for Armano et al.~\cite{Armano}, i.e. the LISA Pathfinder mission to explore in-space detection of gravitational waves. In this experiment a displacement is measured by means of a laser interferometric arm.  However, this concerns a differential acceleration noise measurement and is in that sense a relative measurement instead of an absolute measurement of the acceleration noise. This may have implications for the applicability of these experimental results to bound the models. Furthermore, in their derivation, Oppenheim et al. note that the bound on $D_2$ will depend on the functional choice of $D_2(\Phi)$ on the Newtonian potential $\Phi$, for which they assume that of a large background potential, i.e. that of the Earth's. However, the LISA Pathfinder measurement is done very far away from the Earth, at its Lagrange point $\mathrm{L_1}$. This may affect the validity of this $\FOM$ calculation.\\

 In terms of $\FOM$, atom interferometry experiments outperform all other types of experiments: Asenbaum et al.~\cite{Assenbaum} improve the $\FOM$ by 25 orders of magnitude in comparison to the Cavendish experiment put forward in the original paper by Oppenheim et al. and the $\FOM$ even drops below the lower bound of the ultra-local discrete model describing the spacetime diffusion. However, the value is still one order of magnitude larger than the limit set by the non-local continuous model. We note that this experiment is qualitatively different from the Cavendish torsion balance experiments in the sense that a phase shift is measured instead of a displacement. This is a relative measurement of the acceleration noise and as such the same considerations apply as for the experiment by Armano et al.~\cite{Armano}. Lastly, although an acceleration sensitivity is measured in the atom interferometer, we are unsure to which extent its results are valid to calculate the upper bound. After all, these atom interferometry experiments closely resemble the matter-wave interferometry experiment that Oppenheim et al. use to set a lower bound based on decoherence experiments \cite{Gerlich}. It raises the question if one and the same experiment can put both an upper and a lower bound on $D_2$.\\

We point out that several low-force noise measurements on Earth can still be improved, such as magnets levitating in a superconducting well making use of the Meissner effect, reported by Fuchs et al.~\cite{Fuchs} and Timberlake et al.~\cite{Timberlake23}. These experiments measure force sensitivity by means of yet another technique, namely Superconducting Quantum Interference Devices (SQUID) detection. This type of experiment is currently not limited by thermal noise. Further improvements in vibration isolation will help to close the gap to the thermal noise floor, potentially outperforming the LIGO mirror.\\

\section{Discussion}
Summarizing, a broad range of experiments lay a solid foundation for a new upper bound to the spacetime diffusion parameter $D_2$ that is at least 15 orders of magnitude better than the first bound provided in the original paper. Experiments for which the applicability of the results to this upper bound calculation are questioned, are indicated with an open marker in Figure \ref{Figure1}. Inclusion of these experiments can improve the upper bound by up to 25 orders of magnitude, thus dropping below the bound set by the ultra-local discrete model. Furthermore, we would like to point out that, using the experimental results of Asenbaum et al., there is still a gap of one order of magnitude (out of 26 orders of magnitude) that needs to be closed for the non-local continuous model.\\

Additionally, it is important to realise that we have only considered an improvement of the $\it{upper}$ bound on the spacetime diffusion parameter in this work. Oppenheim et al. calculate the $\it{lower}$ limit on $D_2$, set by coherence measurements of large-mass quantum superposition states, from interferometry experiments showing coherent quantum interference of fullerene molecules~\cite{Gerlich}. Although we have not elaborately reviewed experimental results of coherence measurements, we argue that this lower bound may be raised by examining more novel experiments. For instance, Fein et al.~\cite{Fein} present an updated result of the matter-wave interferometry experiment, which Oppenheim et al. put forward as a first lower bound, where the mass in superposition now exceeds \SI{25}{kDa} and the decoherence rate is given by $\lambda = \tau^{-1}$ = \SI{133}{Hz}. Next, Wang et al.~\cite{Wang} report the coherence of a $^{171}$Yb$^{+}$ ion qubit with a decoherence rate of $\lambda = \SI{182}{\micro Hz}$. However, the superposed state in question is a spin superposition and thus not a mass displacement superposition. Lastly, Bild et al.~\cite{Bild} show a cat state of a $\SI{16}{\micro \gram}$ mechanical oscillator with a decoherence rate of $\lambda \sim \SI{1}{MHz}$. Closer examination of these decoherence experiments may result in new insights on the lower bound on $D_2$.\\

In conclusion, by examining a broad range of modern-day experiments over a large test mass range we can lower the upper bound on the spacetime diffusion parameter $D_2$. If only direct measurements of acceleration sensitivity on Earth are considered, i.e. excluding the open data points in Figure $\ref{Figure1}$, the updated bounds become

\begin{equation}
    \SI{e{-16}}{} \geq \frac{l_P^3}{m_P} D_2 \geq \SI{e{-25}}{}
\end{equation}
for the ultra-local discrete model and

\begin{equation}
    \SI{e{-24}}{} \geq l_P^2 D_2 \geq \SI{e{-35}}{}
\end{equation}
for the non-local continuous model. Here we report a conservative bound. More in-depth scrutiny of the theory put forward by Oppenheim et al. should provide a better understanding of the applicability of the atom interferometry and LISA Pathfinder experimental results.\\

In order to discuss future improvements of $\FOM$, the origin of the force noise in the experiments in Figure \ref{Figure1} is considered. We distinguish between thermal force noise $S_F^{\textrm{th}}$ and other noise sources. The former can be expressed as

\begin{equation}
    S_F^{\textrm{th}} = 4 k_{\textrm{B}} T m \gamma = 4 k_{\textrm{B}} T m \omega_0/Q
    \label{thermal}
\end{equation}

\noindent with $T$ the mode temperature and $\gamma$ the damping coefficient of the oscillating mode of the test mass. Furthermore, the resonance frequency is denoted as $\omega_0$ and the quality factor is denoted as $Q$ for the oscillating mode. An experiment is considered to be thermally limited if the thermal force noise $S_F^{\textrm{th}}$ is larger than half the measured total force noise $S_F$. This distinction is made since thermal force noise has a lower limit set by the fluctuation-dissipation theorem. The $\FOM$ calculated from only the thermal force noise of an experiment is plotted in Figure~\ref{Figure1} if it is at most a factor of two lower than the $\FOM$ calculated from the measured total force noise. This is the case for at least four experiments for which the force or acceleration sensitivity is measured directly. Other sources of force noise generally originate from imperfections in the measurement setup, such as insufficient vibration isolation.\\


Assuming an experiment is thermally limited, we have $\FOM = 4Nk_B T \omega_0/(mQ) \sim T \omega_0 / Q$. This relation shows that lowering the test mass $m$ is not necessary to improve the $\FOM$, as also becomes clear from Figure \ref{Figure1}. However, this relation indicates that the parameters $T$, $\omega_0$ and $Q$ can be used to lower the thermal noise floor. Lowering the temperature $T$ of an experiment often requires drastic changes to the experimental setup, such as carrying out the experiment inside a dilution refrigerator or using vibration isolation systems~\cite{DeWitVibIso}. The parameter $Q$ is related to the amount of damping in a mechanical oscillator. Improving the $\FOM$ in terms of damping would require improved fabrication of the, often nanomechanical, oscillator. This is generally the preferred direction for improving force-sensitivity experiments, as is reflected in the recent development of a broad range of high-$Q$ resonators \cite{EichlerReview}. In terms of the resonance frequency $\omega_0$ we observe for the reviewed experiments that the $\FOM$ is slightly better for experiments with a lower $\omega_0$. Lowering  the resonance frequency $\omega_0$ of an experiment also comes with technical challenges, as vibration isolation at ever lower frequencies becomes increasingly difficult. \\

Summarizing, several of the presented experiments pave the way to further limit the bounds they put on the spacetime diffusion parameter. We look forward to further discussions on the validity of the upper bounds provided by the atom interferometry and LISA Pathfinder experiments, as well as other ways to evaluate the theory for the spacetime diffusion parameter put forward by Oppenheim et al. \\

\section*{Acknowledgments}
We thank P. Logman and G. van de Stolpe for their contributions to the manuscript. This work was supported by the European Union (ERC StG, CLOSEtoQG, Project 101041115), the EU Horizon Europe EIC Pathfinder project QuCoM (10032223) and the NWO grant OCENW.GROOT.2019.088.

\section*{Appendix A: overview of experiments}
\label{appendixA}
For this paper we have reviewed a broad range of different experimental platforms that aim to measure force or acceleration with ultra-high sensitivity. These include (mechanical) resonators with high quality factors that for instance make use of soft clamping or levitation. Other types of experiments are more mesoscopic or macroscopic (massive), such as modern-day torsion pendula and the latest vibration-isolated platforms employed by the LIGO and LISA Pathfinder collaboration. Yet another type of experiment that we have included, are atom interferometry measurements. These are qualitatively different in the sense that they measure a phase shift instead of a displacement, but provide nonetheless very precise acceleration measurements.\\

We have included measurements of force or acceleration sensitivity that are actual experimental results (as opposed to theoretically achievable values). These include for example measurements with SQUIDs or laser interferometry. We have not included papers that only report experimental results of e.g. displacement of strain sensitivity measurements (as opposed to force or acceleration sensitivity), although we think that these measurements could also give potentially interesting results for $\FOM$. In order to find relevant papers, we thankfully consulted previous reviews on precision force-sensitivity measurements~\cite{EichlerReview, CarneyReview, DeBonis, Liang, MooreReview}.\\

Based on test mass magnitude, fabrication technique and measurement method, we distinguish ten different types of experiments. They are clustered as follows:

\begin{itemize}
    \item suspended systems: (carbon) nanotubes, nanowires (e.g. suspsended SiC or Al wires), nanobeams (e.g. GaAs or single-cryslal Si cantilevers), SiN membranes, mesoscopic (e.g. milligram torsion pendulum) and massive (e.g. the LIGO mirror),
    \item levitated systems: trapped ion, optical levitation (e.g. silica nanobeads in optical tweezers) and magnetic levitation (e.g. magnets in superconducting wells),
    \item free-falling systems: atom interferometry.
\end{itemize}

For every category, the two or three experiments that gave the best $\FOM$ were plotted in Figure \ref{Figure1}. To calculate $\FOM$, we have looked up the element and mass of the test mass, from which we could derive the number of nuclei. Additionally, we listed the spectral density of the force and acceleration sensitivity. These values are also listed in Table I. If available in the paper and applicable to the experiment, we have also looked up the resonance frequency, quality factors, mode temperatures and damping coefficient in order to be able to calculate the thermal noise floor via Equation \ref{thermal}.\\

\begin{table}[ht] \label{tab:experiments_1}
\tiny
\centering
\caption{Overview of all experiments that were considered in this paper. Note that not all experiments presented in this table are plotted in Figure \ref{Figure1}. For each experiment the measured parameters relevant for calculating the figure of merit, $\FOM$, are shown. The columns in this table contain from left to right: a reference to the original paper, the type of experiment, the material of the mechanical oscillator, the test mass $m$, the number of nuclei $N$, the resonance frequency $f_0$, the total force noise $S_F$, the acceleration noise $S_a$ and the figure of merit $\FOM$ of the experiment. ($^{\dagger}$) If a value in the table was unavailable in the original paper, we report the value mentioned in the review paper. ($^{*}$) The material described is borosilicate glass. We assumed this to consist for 80$\%$ of SiO$_2$ and for 20$\%$ of B$_2$O$_3$.}
\setlength\tabcolsep{7pt}
\def\arraystretch{2}

\begin{center}

\begin{tabular}{|l|c|c|c|c|c|c|c|c|}

\hline
Reference & Type & Element & m [kg] & N [-] & $f_0$ [Hz] & $\sqrt{S_F}$ [N/$\sqrt{\si{\hertz}}$] & $\sqrt{S_a}$ [ms$^{-2}/\sqrt{\si{\hertz}}$] & $\FOM$ [$m^2s^{-3}$]\\   
\hline
\hline
Asenbaum '17 \cite{Assenbaum} & Atom interfer. & Rb & 1.44 $\cdot 10^{-19}$ & 1.00 $\cdot 10^{6}$ & - & 7.08 $\cdot 10^{-28}$ & 4.91 $\cdot 10^{-9}$ & 2.41 $\cdot 10^{-11}$ \\
Biedermann '15 \cite{Biedermann} & Atom interfer. & Cs & 2.21 $\cdot 10^{-17}$ & 1.00 $\cdot 10^{8}$ & - & 9.09 $\cdot 10^{-25}$ & 4.12 $\cdot 10^{-8}$ & 1.70 $\cdot 10^{-7}$ \\
Armano '18 \cite{Armano} & Massive & Au & 1.93 & 5.89 $\cdot 10^{24}$ & - & 3.35 $\cdot 10^{-15}$ & 1.74 $\cdot 10^{-15}$ & 1.78 $\cdot 10^{-5}$ \\
Gisler '22 \cite{Gisler} & Nanowire & Si$_3$N$_4$ & 9.30 $\cdot 10^{-15}$ & 2.79 $\cdot 10^{11}$ & 1.41 $\cdot 10^{6}$ & 9.60 $\cdot 10^{-21}$ & 1.03 $\cdot 10^{-6}$ & 2.98 $\cdot 10^{-1}$ \\
Seis '22 \cite{Seis} & Membrane & Si$_3$N$_4$ & 1.50 $\cdot 10^{-11}$ & 4.51 $\cdot 10^{14}$ & 1.49 $\cdot 10^{6}$ & 6.50 $\cdot 10^{-19}$ & 4.33 $\cdot 10^{-8}$ & 8.46 $\cdot 10^{-1}$ \\
Martynov '16 \cite{Martynov, CarneyReview}$^{\dagger}$ & Massive & SiO$_2$ & 4.00 $\cdot 10^{1}$ & 1.20 $\cdot 10^{27}$ & 1.00 $\cdot 10^{2}$ & 1.57 $\cdot 10^{-12}$ & 3.92 $\cdot 10^{-14}$ & 1.85 \\
Fogliano '21 \cite{Fogliano} & Nanowire & SiC & 1.60 $\cdot 10^{-14}$ & 4.82 $\cdot 10^{11}$ & 1.16 $\cdot 10^{4}$ & 4.00 $\cdot 10^{-20}$ & 2.50 $\cdot 10^{-6}$ & 3.01 \\
Fuchs '24 \cite{Fuchs} & Magnetic lev.  & Nd$_2$Fe$_{14}$B & 4.00 $\cdot 10^{-7}$ & 3.79 $\cdot 10^{18}$ & 2.67 $\cdot 10^{1}$ & 5.00 $\cdot 10^{-16}$ & 1.25 $\cdot 10^{-9}$ & 5.92 \\
Hamilton '15 \cite{Hamilton} & Atom interfer. & Cs & 4.41 $\cdot 10^{-18}$ & 2.00 $\cdot 10^{7}$ & - & 2.60 $\cdot 10^{-21}$ & 5.89 $\cdot 10^{-4}$ & 6.93 \\
Mamin '01 \cite{Mamin} & Nanobeam & Si & 6.85 $\cdot 10^{-13}$ & 1.47 $\cdot 10^{13}$ & 4.98 $\cdot 10^{3}$ & 8.20 $\cdot 10^{-19}$ & 1.20 $\cdot 10^{-6}$ & 2.11 $\cdot 10^{1}$ \\
Timberlake '23 \cite{Timberlake23} & Magnetic lev. & Nd$_2$Fe$_{14}$B & 2.30 $\cdot 10^{-8}$ & 2.18 $\cdot 10^{17}$ & 4.24 $\cdot 10^{1}$ & 3.40 $\cdot 10^{-16}$ & 1.48 $\cdot 10^{-8}$ & 4.76 $\cdot 10^{1}$ \\
Liang '22 \cite{Liang} & Optical lev. & SiO$_2$ & 4.68 $\cdot 10^{-18}$ & 1.41 $\cdot 10^{8}$ & 1.75 $\cdot 10^{5}$ & 4.34 $\cdot 10^{-21}$ & 9.27 $\cdot 10^{-4}$ & 1.21 $\cdot 10^{2}$ \\
Maiwald '09 \cite{Maiwald} & Trapped ion & Mg$^{+}$ & 4.04 $\cdot 10^{-26}$ & 1.00 & 1.00 $\cdot 10^{6}$ & 4.60 $\cdot 10^{-25}$ & 1.14 $\cdot 10^{1}$ & 1.30 $\cdot 10^{2}$ \\
Norte '16 \cite{Norte} & Membrane & Si$_3$N$_4$ & 2.30 $\cdot 10^{-11}$ & 6.91 $\cdot 10^{14}$ & 1.50 $\cdot 10^{5}$ & 1.00 $\cdot 10^{-17}$ & 4.35 $\cdot 10^{-7}$ & 1.31 $\cdot 10^{2}$ \\
Monteiro '20 \cite{Monteiro20} & Optical lev. & SiO$_2$ & 9.42 $\cdot 10^{-13}$ & 2.27 $\cdot 10^{14}$ & 6.30 $\cdot 10^{1}$ & 8.78 $\cdot 10^{-19}$ & 9.32 $\cdot 10^{-7}$ & 1.97 $\cdot 10^{2}$ \\
Kampel '17 \cite{Kampel, CarneyReview}$^{\dagger}$ & Membrane & Si$_3$N$_4$ & 1.00 $\cdot 10^{-11}$ & 3.00 $\cdot 10^{14}$ & 1.60 $\cdot 10^{6}$ & 9.81 $\cdot 10^{-18}$ & 9.81 $\cdot 10^{-7}$ & 2.89 $\cdot 10^{2}$ \\
Héritier '18 \cite{Heritier} & Nanowire & C & 4.10 $\cdot 10^{-15}$ & 2.06 $\cdot 10^{11}$ & 2.50 $\cdot 10^{4}$ & 1.88 $\cdot 10^{-19}$ & 4.59 $\cdot 10^{-5}$ & 4.33 $\cdot 10^{2}$ \\
Tebbenjohanns '20 \cite{Tebbenjohanns20, Liang}$^{\dagger}$ & Optical lev. & SiO$_2$ & 3.49 $\cdot 10^{-18}$ & 1.05 $\cdot 10^{8}$ & 1.46 $\cdot 10^{5}$ & 8.00 $\cdot 10^{-21}$ & 2.29 $\cdot 10^{-3}$ & 5.52 $\cdot 10^{2}$ \\
Tebbenjohanns '19 \cite{Tebbenjohanns19, Liang}$^{\dagger}$ & Optical lev. & SiO$_2$ & 3.49 $\cdot 10^{-18}$ & 1.05 $\cdot 10^{8}$ & 1.46 $\cdot 10^{5}$ & 1.00 $\cdot 10^{-20}$ & 2.87 $\cdot 10^{-3}$ & 8.63 $\cdot 10^{2}$ \\
Lewandowski '21 \cite{Lewandowski} & Magnetic lev. & SiO$_2$, B$_2$O$_3$ & 2.50 $\cdot 10^{-10}$ & 8.26 $\cdot 10^{15}$ & 1.75 & 8.83 $\cdot 10^{-17}$ & 3.53 $\cdot 10^{-7}$ & 1.03 $\cdot 10^{3}$ \\
Teufel '09 \cite{Teufel} & Nanowire & Al & 5.50 $\cdot 10^{-15}$ & 1.23 $\cdot 10^{11}$ & 1.04 $\cdot 10^{6}$ & 5.10 $\cdot 10^{-19}$ & 9.27 $\cdot 10^{-5}$ & 1.05 $\cdot 10^{3}$ \\
Cripe '19 \cite{Cripe, CarneyReview}$^{\dagger}$ & Nanobeam & GaAs & 5.00 $\cdot 10^{-11}$ & 4.16 $\cdot 10^{14}$ & 8.76 $\cdot 10^{2}$ & 9.81 $\cdot 10^{-17}$ & 1.96 $\cdot 10^{-6}$ & 1.60 $\cdot 10^{3}$ \\
Reinhardt '16 \cite{Reinhardt} & Membrane & Si$_3$N$_4$ & 4.00 $\cdot 10^{-12}$ & 1.20 $\cdot 10^{14}$ & 4.08 $\cdot 10^{4}$ & 2.00 $\cdot 10^{-17}$ & 5.00 $\cdot 10^{-6}$ & 3.00 $\cdot 10^{3}$ \\
Gieseler '13 \cite{Gieseler} & Optical lev. & SiO$_2$ & 3.00 $\cdot 10^{-18}$ & 9.03 $\cdot 10^{7}$ & 1.25 $\cdot 10^{5}$ & 2.00 $\cdot 10^{-20}$ & 6.67 $\cdot 10^{-3}$ & 4.01 $\cdot 10^{3}$ \\
Deli\'{c} '20 \cite{Delic} & Optical lev. & SiO$_2$ & 2.83 $\cdot 10^{-18}$ & 8.52 $\cdot 10^{7}$ & 3.05 $\cdot 10^{5}$ & 1.94 $\cdot 10^{-20}$ & 6.87 $\cdot 10^{-3}$ & 4.02 $\cdot 10^{3}$ \\
Corbitt '07 \cite{Corbitt} & Mesoscopic & SiO$_2$ & 1.00 $\cdot 10^{-3}$ & 3.01 $\cdot 10^{22}$ & 1.80 $\cdot 10^{3}$ & 3.95 $\cdot 10^{-13}$ & 3.95 $\cdot 10^{-10}$ & 4.70 $\cdot 10^{3}$ \\
Westphal '21 \cite{Westphal} & Mesoscopic & Au & 2.18 $\cdot 10^{-4}$ & 6.66 $\cdot 10^{20}$ & 3.59 $\cdot 10^{-3}$ & 9.07 $\cdot 10^{-13}$ & 4.16 $\cdot 10^{-9}$ & 1.15 $\cdot 10^{4}$ \\
Rider '18 \cite{Rider} & Optical lev. & SiO$_2$ & 1.53 $\cdot 10^{-13}$ & 4.62 $\cdot 10^{12}$ & 2.50 $\cdot 10^{2}$ & 1.15 $\cdot 10^{-17}$ & 7.50 $\cdot 10^{-5}$ & 2.60 $\cdot 10^{4}$ \\
Kawasaki '20 \cite{Kawasaki} & Optical lev. & SiO$_2$ & 8.40 $\cdot 10^{-14}$ & 2.53 $\cdot 10^{12}$ & 3.01 $\cdot 10^{2}$ & 1.00 $\cdot 10^{-17}$ & 1.19 $\cdot 10^{-4}$ & 3.58 $\cdot 10^{4}$ \\
Hempston '17 \cite{Hempston} & Optical lev. & SiO$_2$ & 7.60 $\cdot 10^{-19}$ & 2.29 $\cdot 10^{7}$ & 7.20 $\cdot 10^{4}$ & 3.20 $\cdot 10^{-20}$ & 4.21 $\cdot 10^{-2}$ & 4.06 $\cdot 10^{4}$ \\
De Bonis '18 \cite{DeBonis} & Nanotube & C & 8.60 $\cdot 10^{-21}$ & 4.32 $\cdot 10^{5}$ & 2.92 $\cdot 10^{7}$ & 4.30 $\cdot 10^{-21}$ & 5.00 $\cdot 10^{-1}$ & 1.08 $\cdot 10^{5}$ \\
Hälg '21 \cite{Halg} & Membrane & Si$_3$N$_4$ & 1.40 $\cdot 10^{-11}$ & 4.21 $\cdot 10^{14}$ & 1.42 $\cdot 10^{6}$ & 2.80 $\cdot 10^{-16}$  & 2.00 $\cdot 10^{-5}$ & 1.68 $\cdot 10^{5}$ \\
Priel '22 \cite{Priel} & Optical lev. & SiO$_2$ & 5.85 $\cdot 10^{-13}$ & 1.76 $\cdot 10^{13}$ & 1.00 $\cdot 10^{5}$ & 1.00 $\cdot 10^{-16}$ & 1.71 $\cdot 10^{-4}$ & 5.14 $\cdot 10^{5}$ \\
Moser '13 \cite{Moser} & Nanotube & C & 1.00 $\cdot 10^{-20}$ & 5.02 $\cdot 10^{5}$ & 4.20 $\cdot 10^{6}$ & 1.20 $\cdot 10^{-20}$ & 1.20 & 7.23 $\cdot 10^{5}$ \\
Weber '16 \cite{Weber} & Nanotube & C & 9.60 $\cdot 10^{-18}$ & 4.82 $\cdot 10^{8}$ & 4.60 $\cdot 10^{7}$ & 3.90 $\cdot 10^{-19}$ & 4.06 $\cdot 10^{-2}$ & 7.95 $\cdot 10^{5}$ \\
Ranjit '16 \cite{Ranjit16} & Optical lev. & SiO$_2$ & 3.75 $\cdot 10^{-17}$ & 1.13 $\cdot 10^{9}$ & 2.83 $\cdot 10^{3}$ & 1.63 $\cdot 10^{-18}$ & 4.35 $\cdot 10^{-2}$ & 2.14 $\cdot 10^{6}$ \\
Krause '12 \cite{Krause} & Membrane & Si$_3$N$_4$ & 1.00 $\cdot 10^{-11}$ & 3.00 $\cdot 10^{14}$ & 2.75 $\cdot 10^{4}$ & 9.81 $\cdot 10^{-16}$ & 9.81 $\cdot 10^{-5}$ & 2.89 $\cdot 10^{6}$ \\
Nichol '12 \cite{Nichol} & Nanowire & Si & 2.66 $\cdot 10^{-17}$ & 5.72 $\cdot 10^{8}$ & 7.86 $\cdot 10^{5}$ & 1.95 $\cdot 10^{-18}$ & 7.33 $\cdot 10^{-2}$ & 3.07 $\cdot 10^{6}$ \\
Biercuk '10 \cite{Biercuk} & Trapped ion & Be$^{+}$ & 1.95 $\cdot 10^{-24}$ & 1.30 $\cdot 10^{2}$ & 8.67 $\cdot 10^{5}$ & 3.90 $\cdot 10^{-22}$ & 2.00 $\cdot 10^{2}$ & 5.22 $\cdot 10^{6}$ \\
Ranjit '15 \cite{Ranjit15} & Optical lev. & SiO$_2$ & 3.75 $\cdot 10^{-14}$ & 1.13 $\cdot 10^{12}$ & 1.07 $\cdot 10^{3}$ & 2.17 $\cdot 10^{-16}$ & 5.79 $\cdot 10^{-3}$ & 3.78 $\cdot 10^{7}$ \\
Affolter '20 \cite{Affolter} & Trapped ion & Be$^{+}$ & 1.50 $\cdot 10^{-24}$ & 1.00 $\cdot 10^{2}$ & 1.58 $\cdot 10^{6}$ & 1.20 $\cdot 10^{-21}$ & 8.02 $\cdot 10^{2}$ & 6.43 $\cdot 10^{7}$ \\
Timberlake '19 \cite{Timberlake19} & Magnetic lev. & Nd$_2$Fe$_{14}$B & 4.00 $\cdot 10^{-6}$ & 3.79 $\cdot 10^{19}$ & 1.94 $\cdot 10^{1}$ & 7.85 $\cdot 10^{-12}$ & 1.96 $\cdot 10^{-6}$ & 1.46 $\cdot 10^{8}$ \\
Guzmán C. '14 \cite{GuzmanCervantes} & Mesoscopic & SiO$_2$ & 2.50 $\cdot 10^{-5}$ & 7.53 $\cdot 10^{20}$ & 1.07 $\cdot 10^{4}$ & 2.45 $\cdot 10^{-11}$ & 9.81 $\cdot 10^{-7}$ & 7.24 $\cdot 10^{8}$ \\
Shaniv '17 \cite{Shaniv} & Trapped ion & Sr$^{+}$ & 1.44 $\cdot 10^{-25}$ & 1.00 & 1.13 $\cdot 10^{6}$ & 2.80 $\cdot 10^{-20}$ & 1.94 $\cdot 10^{5}$ & 3.78 $\cdot 10^{10}$ \\
Blums '18 \cite{Blums} & Trapped ion & Yb$^{+}$ & 2.89 $\cdot 10^{-25}$ & 1.00 & 8.29 $\cdot 10^{5}$ & 3.47 $\cdot 10^{-19}$ & 1.20 $\cdot 10^{6}$ & 1.44 $\cdot 10^{12}$ \\
Cavendish 1798 \cite{Cavendish} & Massive & Pb & 1.00 & 1.00 $\cdot 10^{26}$ & 2.00 $\cdot 10^{-3}$ & 1.00 $\cdot 10^{-6}$ & 1.00 $\cdot 10^{-6}$ & 1.00 $\cdot 10^{14}$\\
\hline
    \end{tabular}
\end{center}
\end{table}

\section*{Bibliography}
\bibliography{bare-references}

\end{document}